\documentclass[
reprint,
superscriptaddress,
showpacs,
preprintnumbers,
nofootinbib,
nobibnotes,
amsmath,
amssymb, 
aps,
prd,
floatfix
]{revtex4-1}

\usepackage{xcolor}

\usepackage[normalem]{ulem}

\usepackage{fontawesome5}

\usepackage[export]{adjustbox}
\usepackage[colorlinks=true,breaklinks=true]{hyperref}
\hypersetup{urlcolor=blue,
	    citecolor=black}

\usepackage{graphicx}% Include figure files
\usepackage{dcolumn}% Align table columns on decimal point
\usepackage{bm}% bold math
\usepackage{amsmath, amssymb}
\usepackage{graphicx}
\usepackage{slashed}
\usepackage{multirow}
\usepackage{float}
\usepackage[usenames,dvipsnames,svgnames]{xcolor}
\usepackage{physics}
\usepackage{soul}

\begin{document}

\preprint{FERMILAB-PUB-26-0099-T}

\title{Open system approach to neutrinos propagating in an ultralight scalar background}

\author{Lua F. T. Airoldi}
\email{lua.airoldi@usp.br}
\affiliation{Instituto de F\'isica, Universidade de S\~ao Paulo, C.P. 66.318, 05315-970 S\~ao Paulo, Brazil}
\affiliation{Theoretical Physics Department, Fermilab, P.O. Box 500, Batavia, IL 60510, USA}

\author{Gustavo F. S. Alves}%
\email{gustavo.figueiredo.alves@usp.br}
\affiliation{Theoretical Physics Department, Fermilab, P.O. Box 500, Batavia, IL 60510, USA}
\affiliation{Northwestern~University,~Evanston,~IL~60208,~USA}

\author{Pedro A. N. Machado}
\email{pmachado@fnal.gov}
\affiliation{Theoretical Physics Department, Fermilab, P.O. Box 500, Batavia, IL 60510, USA}

\author{Peter Vander Griend}
\email{peter.vandergriend@uky.edu}
\affiliation{Theoretical Physics Department, Fermilab, P.O. Box 500, Batavia, IL 60510, USA}
\affiliation{University of Kentucky, Department of Physics and Astronomy, Lexington, KY 40506 USA}

\date{\today}

\begin{abstract}

We examine decoherence in neutrino oscillations induced by an ultralight scalar field coupled to neutrinos. The scalar induces time- and position-dependent shifts in the neutrino mass matrix. Neutrinos sample different field configurations throughout an experimental data-taking period, which leads to damping effects in the oscillation pattern in the form of decoherence. By recasting the neutrino-scalar dynamics within the open quantum systems framework, we establish a mapping between a complete model and phenomenological decoherence approaches. We find that the parameter driving decoherence scales as $L^2/E^2$, where $L$ is the baseline and $E$ is the neutrino energy, as opposed to $L/E$ typically assumed in phenomenological studies of open system approaches to neutrino oscillations.

\end{abstract}

\maketitle

%\tableofcontents

\section{Introduction}
\label{sec:intro}

Nonzero neutrino masses enable a natural interferometer through the oscillation effect~\cite{Kamiokande-II:1989hkh, Kamiokande-II:1992hns, Becker-Szendy:1992ory, GALLEX:1992gcp, Kamiokande:1994sgx, Kamiokande:1996qmi, Allison:1996yb, Super-Kamiokande:1998kpq, Super-Kamiokande:1998qwk, GALLEX:1998kcz, SAGE:1999uje, GNO:2000avz, SNO:2001kpb, Super-Kamiokande:2001ljr, KamLAND:2002uet, K2K:2004iot, KamLAND:2004mhv, Super-Kamiokande:2006yyp, K2K:2006yov, MINOS:2008kxu, DoubleChooz:2011ymz, MINOS:2011amj, T2K:2011ypd, DayaBay:2012fng, RENO:2012mkc}. 
As neutrinos interact only via the weak force, the quantum nature of oscillations can persist and be observed over long distances, from hundreds to thousands of kilometers. 
As such, neutrino oscillations provide a unique probe of physics beyond the Standard Model. 

A useful framework for analyzing modifications of the neutrino oscillation pattern without tying to a specific model is to treat the neutrino system as an open quantum system~\cite{Benatti:2000ph}, where interactions with an external environment affect the development of neutrino oscillations. 
The open system dynamics are typically parametrized through a generic set of operators without specifying the underlying environmental degrees of freedom. 
This strategy was followed by numerous studies on the impact of decoherence on solar, atmospheric, and reactor neutrino experiments, including both existing and future facilities~\cite{Gago:2000qc, Lisi:2000zt, Gago:2002na, Fogli:2003th, Anchordoqui:2005gj, Fogli:2007tx, Farzan:2008zv, Oliveira:2010zzd, deOliveira:2013dia, Bakhti:2015dca, BalieiroGomes:2016ykp, Coelho:2017byq, Coelho:2017zes, Carrasco:2018sca, Carpio:2018gum, BalieiroGomes:2018gtd, Coloma:2018idr, deHolanda:2019tuf, Gomes:2020muc, Stuttard:2021uyw, DeRomeri:2023dht}. 
Recent experimental analyses from IceCube~\cite{ICECUBE:2023gdv} and KM3NeT/ORCA~\cite{KM3NeT:2024jji} also provided constraints for such effects. 
While model-independent approaches are valuable for establishing general constraints on decoherence, relying only on those leaves open the question of what microphysics can actually generate these effects, or in other words, what do we learn by constraining effective parameters?

In this work, we provide a simple realization of neutrino decoherence through the propagation of neutrinos in a scalar ultralight dark matter (ULDM) background.
We first follow the approach typically used in the literature for ULDM coupled to neutrinos: we evolve each neutrino state in a time-modulating scalar background and derive an effective oscillation probability.
Then, we treat the neutrino-scalar dynamics within an open quantum system framework and derive a Lindblad evolution equation by statistically averaging the neutrino evolution over scalar field configurations.
We show explicitly that both approaches result in the same phenomenology and lead to decoherence effects in neutrino oscillations.
We also show that the decoherence effects due to ULDM do not correspond to scenarios typically studied in the literature.

The paper is organized as follows: In Section~\ref{sec:neutrino_scalar}, we introduce the neutrino-ULDM coupling and derive the phase-dependent evolution. Section~\ref{sec:phase_averaging} develops the exact analytical solution through phase averaging. 
In Section~\ref{sec:open_system}, we demonstrate how the open system approach reproduces these results in the appropriate limit, establishing the connection between the two formalisms. 
We discuss the phenomenology of these models in Section~\ref{sec:phenomenology}, and we conclude in Section~\ref{sec:conclusions}.

\section{Neutrinos coupled to an ultralight scalar background}
\label{sec:neutrino_scalar}

Ultralight scalar fields with masses far below the eV scale are well-motivated dark matter candidates~\cite{Dine:1982ah, Hu:2000ke, Wantz:2009it, Hui:2016ltb} (see Ref.~\cite{Ferreira:2020fam} for a review). For this mass range, the occupation number per mode is extremely large, and the field can be well-approximated as a classical, wave-like background~\cite{Suarez:2013iw, Ferreira:2020fam, Hui:2021tkt, Cheong:2024ose}
\begin{equation}
    \phi(\vec x, t) = \phi_0 \cos(m_\phi t - m_\phi\vec{v}_\phi\cdot\vec x + \theta),
    \label{eq:classical_field}
\end{equation}
where $\theta$ is a phase, $\vec v_\phi$ is the dark matter velocity, and $\phi_0 = \sqrt{2\rho_\phi}/m_\phi$ is the field amplitude set by the local energy density $\rho_\phi$, where $\rho_\phi \approx 0.4~\text{GeV}/\text{cm}^3$ if the scalar saturates the local dark matter density.

We assume that this field couples to neutrinos as
\begin{equation}
    \mathcal{L}_{\rm int} = - g_{ij} \phi \bar{\nu}_i \nu_j + \text{h.c.},
    \label{eq:Lint-general}
\end{equation}
where $i,j$ label mass eigenstates and $g_{ij}$ is the coupling constant. 
For concreteness, we assume that $g$ couples diagonally in the mass basis,
\begin{equation}
    \mathcal{L}_{\rm int} = - g_i \phi \bar{\nu}_i \nu_i + \text{h.c.}
    \label{eq:Lint}
\end{equation}
Extensions to off-diagonal couplings are straightforward. 
This interaction arises in type-I seesaw frameworks~\cite{Minkowski:1977sc, Mohapatra:1979ia} where the scalar couples to heavy right-handed neutrinos~\cite{Fardon:2005wc, Krnjaic:2017zlz, Losada:2021bxx, Huang:2022wmz, Dev:2022bae, ChoeJo:2023ffp}. We remain agnostic about the specific UV completion and focus on the low-energy phenomenological consequences.

The effect of such a coupling is to modify neutrino masses, leading to
\begin{align}
    \begin{split}
        E_{i} &= \sqrt{|\vb{p}|^2 + (m_i + g_i \phi(t))^2} \\
        &\approx |\vb{p}| + \frac{m_i^2 + 2g_i m_i \phi(t)}{2|\vb{p}|},
    \label{eq:modified_energy}
    \end{split}
\end{align}
where we drop terms quadratic in $g_i\phi_0$. 
This is valid when $g_i \phi_0 \ll m_i$, which we assume throughout. 
This leads to a modification of the neutrino oscillation frequency
\begin{equation}
    \frac{\Delta m_{ij}^2}{2|\vb{p}|} \to \frac{\Delta m_{ij}^2}{2|\vb{p}|} + \frac{\Delta \Tilde{m}^2_{ij}(t,\varphi)}{2|\vb{p}|},
    \label{eq:modified_frequency}
\end{equation}
where $\Delta \widetilde{m}^2_{ij} = 2(g_i m_i - g_j m_j)\phi(t)$.

We assume that the scalar field oscillates slowly compared to the neutrino propagation time, i.e., $m_\phi L \ll 1$ with $L$ being the distance traveled by the neutrino from production to detection. 
Therefore,  each neutrino experiences an approximately static scalar background during its propagation. 
However, across the experimental data taking period, different neutrinos sample different background values. 

To make this more precise, we rewrite the time dependence of Eq.~\eqref{eq:classical_field} as
\begin{align}\label{eq:phase_decomposition}
    \cos(m_\phi t-m_\phi \vec v_\phi\cdot\vec x + \theta)
    \approx&\cos\left[m_\phi (L+t_0) + \theta\right]\\  
    =&\cos\left(m_\phi L + \xi\right)  \approx  \cos{(\xi)},
    \nonumber
\end{align}
where we drop the sub-leading $m_\phi v_\phi$ term, and $L = t-t_0$ is approximately the baseline, i.e., the neutrino time of flight. 
We also define the effective phase 
\begin{equation}
    \xi \equiv m_\phi t_0 + \theta.
    \label{eq:effective_phase}
\end{equation}
If, over the course of a neutrino oscillation experiment, the production time $t_0$ spans many oscillation periods of the scalar field, $\xi$ is uniformly sampled in the interval $[0, 2\pi)$. 
This limit corresponds to $ L \ll 1/m_\phi \ll  T$, where $T$ is the total running time of an experiment. 
The effect of varying $\xi$ leads to characteristic distortions of the oscillation probability explored in previous works~\cite{Berlin:2016woy, Krnjaic:2017zlz, Dev:2020kgz, Losada:2021bxx, Losada:2022uvr, Dev:2022bae, Losada:2023zap, Lin:2023xyk, Delgadillo:2025wxw}.

%%%%%%%%%%%%%%%%%%%%%%%%%%%%%%%%%%%%%%%%%%%%%%%%%%%%%%%%%%%%%

\section{Exact solution via phase averaging}
\label{sec:phase_averaging}

The total Hamiltonian in the mass basis takes the form
\begin{equation}
    H(\xi) = H_0 + H_\phi(\xi),
    \label{eq:total_H}
\end{equation}
where the vacuum Hamiltonian is 
\begin{equation}
    H_0 = \frac{1}{2E}\hat{m}_\nu^2,
\end{equation}
where $\hat m_\nu={\rm diag}(m_1,m_2,m_3)$ and the scalar-induced contribution is
\begin{equation}\label{eq:H_phi}
    H_\phi(\xi) = \frac{\phi_0}{E}\hat g \hat m_\nu\cos(\xi),
\end{equation}
where $\hat g$ denotes the matrix of coupings $g_{ij}$.
Here both $H_0$ and $H_\phi$ are diagonal in the mass basis and thus they commute. 
This leads to a factorization of the time-evolution operator 
\begin{equation}
    U(t, \xi) = 
    U_0(t)  U_\phi(t,\xi),
    \label{eq:U_factorization}
\end{equation}
where $U_0(t) = \exp(-iH_0 t)$ is the standard vacuum evolution and $U_\phi(t,\xi) = \exp[-iH_\phi(\xi)t]$ encodes the effect of the scalar field; this quantity should be averaged over $\xi$.

In order to do this averaging systematically and in a way that makes comparison to the open system formalism easier, we use the density matrix formalism and introduce the vectorization of the time evolution. 
We start by recalling that any $3 \times 3$ Hermitian matrix can be expanded in terms of the Gell-Mann matrices $\{\lambda_\mu\}$ and the $3\times 3$ identity matrix. 
The neutrino density matrix can be decomposed as
\begin{equation}
    \rho = \sum_{\mu=0}^{8} \rho^\mu \lambda_\mu,
    \label{eq:rho_expansion}
\end{equation}
where $\lambda_0^{ij} = \delta_{ij}/\sqrt{3}$ and $\lambda_{1,\ldots,8}$ are the Gell-Mann matrices normalized according to $\text{Tr}(\lambda_\mu \lambda_\nu) = \delta_{\mu\nu}$. 
The vectorization of the density matrix is simply a mapping of $\rho$ to a 9-component vector
\begin{equation}
    |\rho\rangle = \begin{pmatrix} \rho^0 \\ \rho^1 \\ \vdots \\ \rho^8 \end{pmatrix}.
    \label{eq:rho_vector}
\end{equation}
Note that in this parametrization, $\rho^0$, $\rho^3$ and $\rho^8$ are the vector entries that enter in the diagonal of the $3\times 3$ representation of $\rho$, that is, in the mass basis they can be mapped onto the coefficients of $|\nu_j\rangle\langle \nu_j|$.

In the vectorized representation, the density matrix time evolution
\begin{equation}
    \frac{\partial \rho}{\partial t} = -i[H, \rho]
\end{equation}
takes the form
\begin{equation}
    \frac{\partial}{\partial t}|\rho(t)\rangle = \mathcal{H}|\rho(t)\rangle,
    \label{eq:vectorized_evolution}
\end{equation}
where $\mathcal{H}$ is a $9 \times 9$ matrix. 
The formal solution of the differential equation is
\begin{equation}
    |\rho(t, \xi)\rangle = \mathcal{U}(t, \xi) |\rho(0)\rangle,
    \label{eq:vectorized_solution}
\end{equation}
with $\mathcal{U}(t, \xi) = e^{\mathcal{H}(\xi) t}$.
As the coupling of the scalar is diagonal in the mass basis, the evolution operator can be written explicitly as
\begin{widetext}
    \begin{equation}
        \mathcal{U}(t,\xi) = \begin{pmatrix}
1 & 0 & 0 & 0 & 0 & 0 & 0 & 0 & 0 \\
0 & \cos\Delta_{21}^{\rm eff} & \sin\Delta_{21}^{\rm eff} & 0 & 0 & 0 & 0 & 0 & 0 \\
0 & -\sin\Delta_{21}^{\rm eff} & \cos\Delta_{21}^{\rm eff} & 0 & 0 & 0 & 0 & 0 & 0 \\
0 & 0 & 0 & 1 & 0 & 0 & 0 & 0 & 0 \\
0 & 0 & 0 & 0 & \cos\Delta_{31}^{\rm eff} & \sin\Delta_{31}^{\rm eff} & 0 & 0 & 0 \\
0 & 0 & 0 & 0 & -\sin\Delta_{31}^{\rm eff} & \cos\Delta_{31}^{\rm eff} & 0 & 0 & 0 \\
0 & 0 & 0 & 0 & 0 & 0 & \cos\Delta_{32}^{\rm eff} & \sin\Delta_{32}^{\rm eff} & 0 \\
0 & 0 & 0 & 0 & 0 & 0 & -\sin\Delta_{32}^{\rm eff} & \cos\Delta_{32}^{\rm eff} & 0 \\
0 & 0 & 0 & 0 & 0 & 0 & 0 & 0 & 1
\end{pmatrix},
    \label{eq:vectorized_evol}
    \end{equation}
\end{widetext}
where, setting $t=L$,
\begin{equation}
    \Delta_{ij}^{\rm eff} = \frac{\Delta m^2_{ij}L}{2E} + \Delta_{ij}^{\phi}\cos{\xi},
    \label{eq:effective_phase_ij}
\end{equation}
and
\begin{equation}
    \Delta^\phi_{ij} = \frac{\phi_0 (g_im_i - g_jm_j)L}{E}.
    \label{eq:phi_phase}
\end{equation}
To compare with previous literature~\cite{Krnjaic:2017zlz, Delgadillo:2025wxw}, we can write 
\begin{equation}
    \Delta_{ij}^{\rm eff} = \frac{\Delta m^2_{ij}L}{2E}(1 + 2\eta_\phi^{ij}\cos{\xi}),
    \label{eq:effective_phase_ij}
\end{equation}
where
\begin{equation}\label{eq:eta}
    \eta_\phi^{ij} = \frac{\phi_0 (g_im_i - g_jm_j)}{\Delta m^2_{ij}},
\end{equation}
encodes the fractional effect of the scalar field on the neutrino mass splittings.

\subsection{Phase averaging}

The density matrix describing the neutrino experiment is obtained by averaging over the effective phase 
\begin{align}
    \begin{split}
        \overline{\rho}(t) &= \frac{1}{2\pi} \int_0^{2\pi}\! d\xi \, \rho(t, \xi)\\ 
        &= \frac{1}{2\pi} \int_0^{2\pi}\! d\xi \, U(t, \xi) \rho(0) U^\dagger(t, \xi),
        \label{eq:phase_average}
    \end{split}
\end{align}
or, in the vectorized representation, 
\begin{equation}
    |\overline{\rho}(t)\rangle = \overline{\mathcal{U}}(t) |\rho(0)\rangle = \left(\frac{1}{2\pi} \int_0^{2\pi} d\xi \, \mathcal{U}(t, \xi)\right) |\rho(0)\rangle.
     \label{eq:averaged_superoperator}
\end{equation}

For our evolution operator (\ref{eq:vectorized_evol}), 
it is straightforward to show that its average can be written as
\begin{equation}
    \overline{\mathcal{U}}(t) = \mathcal{U}_0(t)  \mathcal{D}(t),
    \label{eq:averaged_result}
\end{equation}
where $\mathcal{U}_0(t)$ is the standard vacuum evolution operator, given by setting $\Delta_{ij}^\phi = 0$ in  
Eq.~\eqref{eq:vectorized_evol},and $\mathcal{D}(t)$ is given by
\begin{align}
    \begin{split}
        \mathcal{D}(t) = \text{diag}&\left(1, J_0(\Delta_{21}^\phi), J_0(\Delta_{21}^\phi), 1, J_0(\Delta_{31}^\phi),\right.\\
        &\quad\left.J_0(\Delta_{31}^\phi), J_0(\Delta_{32}^\phi), J_0(\Delta_{32}^\phi),    1\right).
    \end{split}
    \label{eq:damping_matrix}
\end{align}
with $J_0$ being the zeroth Bessel function.
We note that the entries of $\mathcal{D}(t)$ acting on $\rho^{0}$, $\rho^{3}$ and $\rho^{8}$ are unity.
Thus only the off-diagonal entries are modified by the phase averaging, leading to decoherence in neutrino oscillations.

We can now compute the oscillation probabilities. 
As it was shown before~\cite{Krnjaic:2017zlz, Delgadillo:2025wxw}, the JUNO experiment~\cite{JUNO:2015zny, JUNO:2021vlw} has the potential to provide strong constraints on ultralight dark matter coupled to neutrinos. 
For simplicity, we neglect matter effects in JUNO (though see Ref.~\cite{Khan:2019doq}). 
The survival probability for an electron antineutrino propagating through an ultralight scalar background is
\begin{align}\label{eq:prob_bessel}
    \overline{P}_{ee}  =& \, 1 - \frac12 c_{13}^4\sin^22\theta_{12} - \frac12 \sin^22\theta_{13} \nonumber \\
    &+ \frac{1}{2}c_{13}^4 \sin^22\theta_{12} J_0^{21}(\Delta^\phi_{21})
    \cos(2\Delta_{21}) \\
    & + \frac{1}{2}\sin^22\theta_{13} \left[c_{12}^2 J_0^{31}(\Delta^\phi_{31}) \cos(2\Delta_{31})\right.\nonumber\\
    &\left.+ s^2_{12} J_0^{32}(\Delta^\phi_{32}) \cos(2\Delta_{32})\right],
    \nonumber
\end{align}
where $s_{ij}\equiv\sin\theta_{ij}$, $c_{ij}\equiv\cos\theta_{ij}$, and $\Delta_{ij}\equiv\Delta m^2_{ij}L/4E$ are the standard three-flavor vacuum oscillation phases. 

To understand the impact of the ultralight field on oscillations, we can expand $J_0^{ij} \approx 1 - (\Delta_{ij}^\phi)^2/4$, recovering standard oscillations with small corrections. 
When $\Delta_{ij}^\phi \gtrsim 1$, the Bessel function oscillates with decreasing amplitude towards zero, shutting off the oscillatory pattern.
The same behavior happens when averaging the neutrino oscillation probability over many oscillations. 

\section{Open system approach to phase averaging}
\label{sec:open_system}

Having derived the exact analytical solution through direct phase averaging in the previous section, we now show how the same physics can be captured within the framework of open quantum systems. 
This connection is valuable for several reasons: it provides an alternative computational approach that extends naturally to more complex scenarios where exact solutions are unavailable and it establishes the precise mapping between phenomenological Lindblad parameters and fundamental model parameters. 
We follow closely Ref.~\cite{Burgess:1996mz}, which studied neutrinos propagating in a fluctuating background from an open system perspective.

We start by decomposing the unitary evolution into
\begin{equation}
    U(t, \xi) = \overline{U}(t) + \Delta U(t, \xi),
    \label{eq:U_decomposition}
\end{equation}
where the overline indicates statistical averaging over the phase $\xi$,
\begin{equation}
    \overline{U}(t) \equiv \frac{1}{2\pi} \int_0^{2\pi} d\xi \, U(t, \xi).
    \label{eq:deltaU_average}
\end{equation}
Note that, by construction, $\overline{\Delta U} = 0$. This decomposition separates the dynamics into two contributions: an average effect and that of fluctuations of the background.

Substituting this decomposition into the phase-averaged density matrix, we have
\begin{align}
        \overline{\rho}(t) &= \frac{1}{2\pi} \int_0^{2\pi} d\xi \, U(t, \xi) \rho(0) U^\dagger(t, \xi) \nonumber \\
        &= \overline{U}(t) \rho(0) \overline{U}^\dagger(t) + \overline{\Delta U(t) \rho(0) \Delta U^\dagger(t)},
        \label{eq:rho_decomposition}
\end{align}
where the cross-terms vanish since $\overline{\Delta U} = 0$. The first term represents the evolution under the averaged dynamics, while the second term captures the contribution from fluctuations and leads to decoherence.

In order to derive a master equation, we expand the evolution operator perturbatively in the coupling $g_i$, see Eqs.~(\ref{eq:total_H}-\ref{eq:H_phi}). 
The evolution operator to second order in the ULDM coupling is
\begin{equation}
    U(t, \xi) \approx 1 - i \int_0^t \!\!d\tau H_\phi(\tau, \xi)\! \left[1+i\! \int_0^\tau \!\!d\tau' H_\phi(\tau', \xi)\right]\!,
    \label{eq:U_expansion}
\end{equation}
where we have defined $H_\phi(\tau, \xi) = e^{iH_0\tau} H_\phi(\xi) e^{-iH_0\tau}$ as the ULDM operator written in the interaction picture.
Note that, since $H_0$ and $H_\phi$ commute, moving to the interaction picture is trivial. The result is
\begin{equation}\label{eq:U_approx}
    U(t,\xi) \approx \mathbf{1} - iH_\phi(\xi)t - \frac{1}{2}H_\phi(\xi)^2 t^2.
\end{equation}
We then use $\Delta U(t,\xi) = U(t,\xi) - \overline U(t)$ and Eq.~\eqref{eq:U_approx} to derive up to order $g$ in the coupling
\begin{equation}
    \Delta U(t, \xi) = 
    -i\frac{\phi_0}{E}\hat g\hat{m}_\nu\cos{(\xi)}\, t.
    \label{eq:deltaU_leading}
\end{equation}
We use this expression to obtain
\begin{equation}
    \overline{\Delta U \rho(0) \Delta U^\dagger} = \frac{ \phi_0^2 t^2}{2E^2}\hat g\hat{m}_\nu \overline{\rho}(0) \hat{m}_\nu \hat g.
\label{eq:fluctuation_term}
\end{equation}
Following Ref.~\cite{Burgess:1996mz}, we introduce an effective potential $V(t)$ defined by
\begin{equation}
    \frac{\partial \overline{U}(t)}{\partial t} = -i V(t) \overline{U}(t).
    \label{eq:effective_potential_def}
\end{equation}
Computing the phase average of Eq.~\eqref{eq:U_approx} and substituting into Eq.~\eqref{eq:effective_potential_def}, it is straightforward to show that up to order $g^{2}$ in the coupling
\begin{equation}
    V(t) = -i\frac{\phi_0^2}{2E^2} \left( \hat{g}\,\hat{m}_{\nu} \right)^2 \, t.
    \label{eq:V_result}
\end{equation}

Taking the time derivative of Eq.~\eqref{eq:rho_decomposition} and moving back to the Schrödinger picture, we write the evolution equation in Lindblad form
\begin{align}
    \frac{\partial \overline{\rho}(t)}{\partial t} = &-i[H_0, \overline{\rho}(t)]\nonumber\\
    &\label{eq:lindblad}
    - \frac{\phi_0^2 t}{2 E^2} \biggr( \{\hat{A}_\nu^\dagger \hat{A}_\nu,\overline{\rho}(t) \} - 2 \hat{A}_\nu \overline{\rho}(t) \hat{A}_\nu^\dagger\biggr),
\end{align}
where $\hat{A}_\nu =\hat{g}\,\hat{m}_\nu$.
We note that this Lindblad equation follows from the substitutions $\rho(0)\to \overline{\rho}(t)$ and $ \overline{U}(t)\rho(0)\overline{U}^{\dagger}(t)\to \overline{\rho}(t)$, which introduce corrections beyond the $g^{2}$ accuracy of the above equation. 

In the vectorized representation, the solution is
\begin{equation}
    \ket{\rho(t)} =  \mathcal{U}_0(t)  \mathcal{D}(t) \ket{\rho(0)},
\end{equation}
where the dissipator $\mathcal{D}$ is 
\begin{equation}
    \mathcal{D} = \text{diag}\left(1, D_{21}, D_{21}, 1, D_{31}, D_{31}, D_{32}, D_{32}, 1\right),
    \label{eq:D_gaussian}
\end{equation}
and
\begin{equation}
    D_{ij} \equiv \exp\left[-\left(\frac{\eta_\phi^{ij}\Delta m_{ij}^2 t}{2 E}\right)^2\right].
    \label{eq:gaussian_damping}
\end{equation}
Note that $t=L$ is the neutrino baseline.

The oscillation probability computed from the open system approach is 
\begin{align}
    \overline{P}_{ee}  =& \, 1 - \frac12 c_{13}^4\sin^22\theta_{12} - \frac12 \sin^22\theta_{13} \nonumber \\
    &+ \frac{1}{2} c_{13}^4 \sin^22\theta_{12} D_{21} 
    \cos(2\Delta_{21}) \\
    & +\frac{1}{2} \sin^22\theta_{13} \left[c_{12}^2 D_{31} \!\cos(2\Delta_{31}) + s^2_{12} D_{32}\! \cos(2\Delta_{32})\right].
    \nonumber
\end{align}
To second order in $g_i$, this probability matches the analytical result in Eq.~\eqref{eq:prob_bessel}.
Therefore, within the weak coupling regime where the open system expansion is valid, the Lindblad treatment reproduces the exact analytical result while providing a mapping from open system descriptions to concrete model parameters.

\section{Phenomenology}\label{sec:phenomenology}

Now that we have derived the Lindblad equation for the decoherence due to ultralight dark matter, we can estimate the experimental sensitivity to this kind of model. 
Note that the distinctive form of the decoherence parameter $\propto L^2/E^2$ differs from the $L/E$ form commonly assumed in the literature~\cite{Gago:2000qc, Lisi:2000zt, Fogli:2003th, Anchordoqui:2005gj, Fogli:2007tx, Farzan:2008zv, Oliveira:2010zzd, deOliveira:2013dia, Bakhti:2015dca, BalieiroGomes:2016ykp, Coelho:2017byq, Coelho:2017zes, Carrasco:2018sca, Carpio:2018gum, BalieiroGomes:2018gtd, Coloma:2018idr, deHolanda:2019tuf, Gomes:2020muc,  DeRomeri:2023dht}; in particular, the $L^2/E^2$ scaling is not included in the latest IceCube search for ``quantum foam''~\cite{ICECUBE:2023gdv} and KM3NeT/ORCA~\cite{KM3NeT:2024jji} experimental constraints. 

The $L^2/E^2$ dependence of the damping effect indicates that experiments which observe the oscillatory pattern for the largest $L/E$ setup are the most sensitive to this effect.
This suggests that JUNO is the most promising experiment to search for these effects, as first proposed in Ref.~\cite{Krnjaic:2017zlz} and recently investigated in  Ref.~\cite{Delgadillo:2025wxw}.
We recast the results of the latter in terms of $\eta_\phi^{ij}$, that is, the fractional change in $\Delta m^2_{ij}$ due to the ULDM, see Eq.~\eqref{eq:eta}.
With a detailed simulation of JUNO, Ref.~\cite{Delgadillo:2025wxw} found $\eta_\phi^{21} \lesssim 2.5\%$ and $\eta_\phi^{31} \lesssim 0.5\%$.

\begin{figure}
    \centering
    \includegraphics[width=\linewidth]{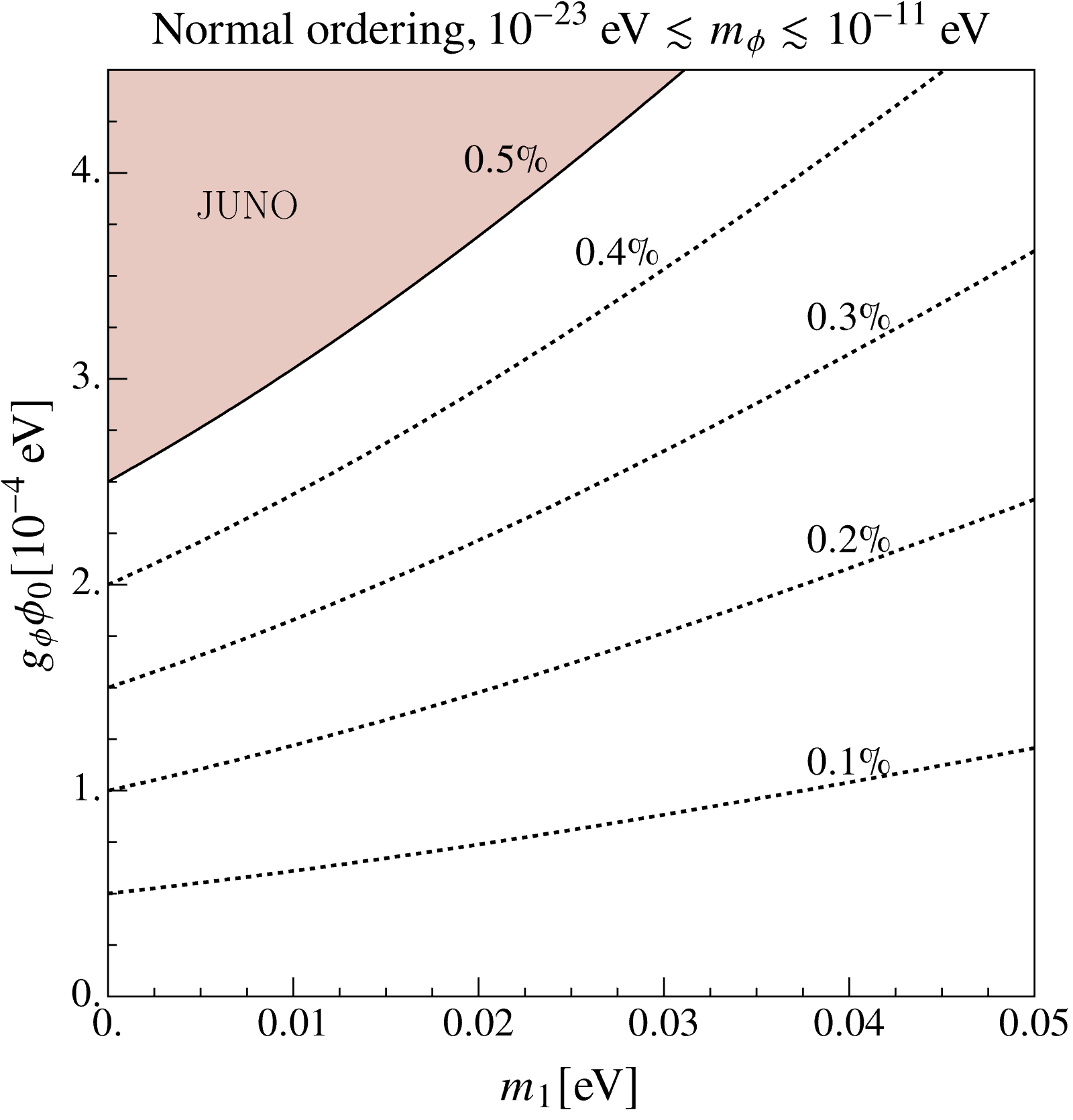}
    \caption{Contours of $\eta_\phi^{31}$ in the plane of the scalar parameters $g \phi_0$ and the lightest neutrino mass $m_1$, assuming normal ordering. The shaded region indicates the parameter space constrained by JUNO at 90\% confidence level~\cite{Delgadillo:2025wxw}.}
    \label{fig:JUNO_bounds}
\end{figure}

In Fig.~\ref{fig:JUNO_bounds}, we show contours of $\eta_\phi^{31}$
as a function of the scalar parameters and the lightest neutrino mass, assuming normal ordering. 
This could correspond to a case in which the scalar couples only to the third neutrino mass eigenstate, i.e., only $g_{3}$ is nonzero.
The JUNO constraint is displayed as the shaded red region. 

We also investigate the sensitivity of DeepCore data~\cite{IceCube:2017lak, IceCube:2019dqi} to distortions of atmospheric neutrino oscillations due to the ultralight scalar field.
Our methods are detailed in App.~\ref{app:IC_analysis}. 
We find that IceCube, for $10^{-23}~{\rm eV}\lesssim m_\phi\lesssim 10^{-14}~{\rm eV}$, can only constrain 
$\eta_\phi^{31} \lesssim 20\%$.~\footnote{IceCube data do not provide a stronger constraint than our starting assumption $\eta_\phi^{21}\ll 1$.} 
Compared to JUNO, these bounds are a factor 40 weaker. 

\subsection{Decoherence, but not necessarily quantum}
\label{sec:decoherence_origin}

It is essential to clarify the physical origin of the decoherence effects derived above, as the terminology can be misleading. 
The decoherence we observe is fundamentally different from microscopic quantum decoherence arising from entanglement with environmental degrees of freedom; see for example Ref.~\cite{Breuer:2002pc}. 
Rather, it emerges from \emph{classical statistical averaging} over an ensemble of neutrinos that sample different field configurations.

This can be trivially understood from the assumption that each neutrino experiences an approximately constant background field configuration, and the entire damping of oscillations comes from the statistical averaging of the field configuration given by the phase $\xi$.
Again, for a neutrino produced at a given time, the background field is approximately constant throughout the propagation of the neutrino.
The apparent decoherence arises only when we consider the \emph{ensemble} of neutrinos detected over the experimental observation period. 
In the open system framework, this is encoded by the ensemble-averaged density matrix
\begin{equation}
    \overline{\rho}(t) = \frac{1}{2\pi}\int_0^{2\pi} d\xi\, \rho_\nu(t,\xi).
\end{equation}
This averaged state is naturally a mixed state ($\text{Tr}[\overline{\rho}^2] < 1$) due to the averaging over a classical probability distribution of initial conditions. 
The Lindblad master equation derived in Section~\ref{sec:open_system} should thus be understood as an effective description of this classical averaging process, as opposed to microscopic quantum decoherence.

In fact, for the parameter space explored, microscopic quantum decoherence is absent. 
To understand this, we must examine the quantum state of the ultralight scalar field itself. 
Microscopic decoherence would arise if the neutrino and ultralight scalar background were entangled throughout the neutrino propagation. 
The decoherence rate would be determined by the two-point correlation function of the environment~\cite{Burgess:1996mz}
\begin{equation}
    C_\phi(x,y) = \text{Tr}_\phi\left[\Delta\phi(x)\Delta\phi(y)\,\rho_\phi\right],
    \label{eq:field_correlator}
\end{equation}
where ${\rm Tr}_{\phi}$ indicates a trace over the environment, i.e., the scalar degrees of freedom; $\Delta\phi = \phi - {\rm Tr}_{\phi}[\phi\rho_\phi]$ represents fluctuations of the field operator around its mean; and $\rho_\phi$ describes the quantum state of the ultralight scalar.

For neutrinos sampling a static or slowly-varying background, the state of the ultralight scalar field can be described by a pure coherent state $\rho_\phi = \ket{\alpha}\bra{\alpha}$, in the sense that $\phi \ket{\alpha}= (e^{-i m t} \alpha + e^{i m t} \alpha^*) \ket{ \alpha}$.
Hence, the correlator vanishes
\begin{equation}
    C_\phi(x,y)= \bra{\alpha}\Delta\phi(x)\Delta\phi(y)\ket{\alpha} = 0.
    \label{eq:coherent_vanishing}
\end{equation}
Since the microscopic decoherence rate is determined by $C_\phi$~\cite{Burgess:1996mz}, the vanishing of this correlator for a pure coherent state implies that no quantum decoherence arises in this regime.

The situation changes if the ultralight scalar is not in a pure coherent state. 
This case corresponds to neutrinos crossing multiple coherent patches of the scalar field.
To illustrate this point, we can write a general density matrix as~\cite{Glauber:1963tx, Sudarshan:1963ts}
\begin{equation}
    \rho_\phi = \int d\alpha\, P(\alpha)\ket{\alpha}\bra{\alpha},
\end{equation}
where $P(\alpha)$ is a quasi-probability distribution over coherent states. 
The characteristic coherence scales that determine over what distances and times the field configuration can be treated as approximately constant are~\cite{Cheong:2024ose}
\begin{align}
    \tau_c &\simeq 2.8~\text{ms}\left(\frac{1 \mu\text{eV}}{m_\phi}\right), \label{eq:coherence_time} \\
    \lambda_c &\sim 674~\text{m}\left(\frac{1 \mu\text{eV}}{m_\phi}\right). \label{eq:coherence_length}
\end{align}
This means that for JUNO, neutrinos would cross different patches for scalars heavier than $10^{-8}$~eV, and microscopic decoherence may become important. 
Note, however, that if the scalar oscillates too fast, the effect may be completely averaged out~\cite{Dev:2020kgz}. 
It would be an interesting exercise to find a scenario where the ultralight scalar background is not described by a pure coherent state throughout the neutrino propagation, while the effects are not averaged out. Such a scenario would leverage neutrinos as a portal to test the actual quantum state of the field. 
We leave this question for future work.

\section{Conclusions}
\label{sec:conclusions}

We have established a complete framework connecting ultralight scalar dark matter to neutrino decoherence, providing a connection to phenomenological open system approaches. 
Starting from the microscopic neutrino-scalar interaction, we derived the exact oscillation probability through phase averaging and demonstrated its equivalence to a Lindblad master equation, establishing a concrete mapping between fundamental parameters $(g, \phi_0, m_\phi)$ and effective decoherence observables.

A key result is the distinctive $L^2/E^2$ scaling of the decoherence parameter, which differs fundamentally from $ L /E$ as typically assumed in experimental searches, as well as more general assumptions of the form $\Gamma L (E/E_0)^n$, with $n$ an integer and $\Gamma$ the damping rate. 
This implies that existing constraints from IceCube and KM3NeT/ORCA do not apply to this class of models. 
The scaling favors experiments with large $L/E$ reach, making JUNO particularly sensitive with projected bounds of $\eta_\phi^{31} \lesssim 0.5\%$. 
Our IceCube analysis yields weaker constraints of $\eta_\phi^{31} \lesssim 19\%$, confirming JUNO's advantage for this scenario.
Importantly, the decoherence observed in our framework arises from classical statistical averaging over different scalar field configurations sampled by neutrinos produced at different times, as opposed to quantum entanglement between the neutrino and the environment. 

Our results demonstrate that model-building can reveal qualitatively different decoherence signatures missed by purely phenomenological approaches.
We hope to motivate targeted experimental searches for $L^2/E^2$ scaling, as well as theoretical investigation of the quantum decoherence regime in neutrino oscillations.

\section{Acknowledgements}
L.A. would like to thank the hospitality of the
Fermilab Theory Division. 
G.F.S.A. and P.M. are supported by Fermi Forward Discovery Group, LLC under Contracts No. 89243024CSC000002 and DE-SC0010143 with the U.S. Department of Energy, Office of Science, Office of High Energy Physics. L.A. 
received full financial support
from the S\~ao Paulo Research Foundation (FAPESP)
through the contracts No. 2025/07427-7 and No. 2024/03749-7.
P.V.G. acknowledges support by the U.S. Department of Energy, Office of
Science, Office of High Energy Physics, under Award DE-SC0019095.
\appendix

\section{IceCube analysis}
\label{app:IC_analysis}
For negative spectral indices, lower neutrino energies provide better sensitivity, making the DeepCore sub-detector particularly suitable. DeepCore is an inner, densely-instrumented region within IceCube with string spacing reduced to 70~m (compared to 125~m in the main array) and digital optical modules vertical separation of only 7~m. This configuration lowers the energy threshold from $E \approx 100$~GeV to $E \approx 5$~GeV, enabling atmospheric neutrino oscillation measurements.

We derive constraints using the publicly available DeepCore atmospheric tau neutrino appearance data~\cite{IceCube:2017lak,IceCube:2019dqi}, collected between April 2012 and May 2015. 
The dataset (Analysis A) consists of an $8 \times 10 \times 2$ dimensional histogram binned in reconstructed energy $E_{\nu}^{\text{reco}}$ (5.6---56~GeV, logarithmically spaced), reconstructed zenith angle $\cos\theta^{\text{reco}}$ (linearly spaced between $-1$ and $1$), and particle identification (PID = 0 for cascade-like events, PID = 1 for track-like events). 
The collaboration provides Monte Carlo simulations including detector systematic uncertainties and event weights accounting for effective area, cross sections, and atmospheric flux.

To compare our decoherence model predictions with the observed data, we perform a $\chi^2$ analysis defined by
\begin{equation}
    \chi^2 = \sum_{i=1}^{\text{bins}} \frac{\left[D_{i} - T_{i}(\vec{\mu})\right]^2}{T_{i}(\vec{\mu})}\,,
\end{equation}
where the sum runs over all bins and $D_{i}$ and $T_{i}(\vec{\mu})$ correspond to the 
number of events in the $i$-th bin for data and theory,
respectively. 
The predicted number of events depends on a parameter set $\vec{\mu}$, defined as follows. 
For the decoherence model, we consider
\begin{equation}
    \mathcal{D} = \text{diag}\left(1, D_{21}, D_{21}, 1, D_{31}, D_{31}, D_{32}, D_{32}, 1\right),
\end{equation}
with
\begin{equation}
    D_{ij} = \exp\left[-\left(\frac{\eta_\phi^{ij}\Delta m_{ij}^2 L}{2 E}\right)^2\right].
\end{equation}
When studying IceCube's sensitivity to $\eta_{\phi}^{21}$, we set $\eta^{31}_\phi = \eta^{32}_\phi = 0$, and for the other scenario, we set $\eta^{31}_\phi=\eta^{32}_\phi$ and $\eta^{21}_\phi=0$. 
We allow the overall normalization of the atmospheric neutrino flux $\mathcal{N}_{\phi_\nu}$ and atmospheric muon flux $\mathcal{N}_{\phi_\mu}$ to float freely in our analysis as nuisance parameters to compensate for cosmic ray flux uncertainties. Therefore, $\vec{\mu} = (\eta_{\phi}^{ij}, \mathcal{N}_{\phi_\nu}, \mathcal{N}_{\phi_\mu})$. We fix the oscillation parameters to their best-fit values from NuFIT~\cite{Esteban:2024eli}. The detector-related systematic uncertainties (optical efficiency, ice scattering, and absorption) are incorporated following the prescription provided by the collaboration and fixed to their best-fit values (Table~II of Ref.~\cite{IceCube:2019dqi}).
We define the $90$\% confidence level upper bounds using the profile likelihood ratio test statistic
\begin{equation}
    \Delta\chi^2 = -2\log\frac{\mathcal{L}(\mu)}{\mathcal{L}(\hat{\mu})}\,,
\end{equation}
where $\hat{\mu}$ is the best-fit value that maximizes the likelihood ($\eta_\phi^{ij} =0$ in our scenario). We find that DeepCore data yield an upper bound $\eta_\phi^{31} \lesssim 20\%$ at $90\%$ confidence level, while for $\eta_\phi^{21}$ the sensitivity is too weak to probe the regime where our assumption $\eta_{\phi}^{ij} \ll 1$ holds.

\bibliography{draft}

\end{document}